\documentclass[a4paper,aps,nofootinbib,floatfix,showpacs,preprintnumbers,twocolumn,superscriptaddress]{revtex4-2} 
\usepackage{graphicx}
\usepackage{bm}
\usepackage{mathrsfs}
\usepackage{float}
\usepackage{tabularx} 
\usepackage{amsmath}
\usepackage{epstopdf}
\usepackage{xcolor}
\usepackage{hyperref}
\usepackage{natbib}
\UseRawInputEncoding
\input epsf
 
\usepackage{amsthm}
\usepackage{mathtools}
\usepackage{physics}
\usepackage{xcolor}
\usepackage{graphicx}
\usepackage[left=23mm,right=13mm,top=35mm,columnsep=15pt]{geometry} 
\usepackage{adjustbox}
\usepackage{placeins}
\usepackage[T1]{fontenc}
\usepackage{lipsum}
\usepackage{csquotes}

\begin{document}

\title{Search for sub-GeV Dark Matter via Migdal effect with an EDELWEISS germanium detector with NbSi TES sensors}

\author{E. Armengaud}
\affiliation{IRFU, CEA, Universit\'{e} Paris-Saclay, F-91191 Gif-sur-Yvette, France}
\author{Q. Arnaud}
\affiliation{Univ Lyon, Universit\'e Lyon 1, CNRS/IN2P3, IP2I-Lyon, F-69622, Villeurbanne, France}
\author{C. Augier}
\affiliation{Univ Lyon, Universit\'e Lyon 1, CNRS/IN2P3, IP2I-Lyon, F-69622, Villeurbanne, France}
\author{A.~Beno\^{i}t}
\affiliation{Institut N\'{e}el, CNRS/UJF, 25 rue des Martyrs, BP 166, 38042 Grenoble, France}
\author{L.~Berg\'{e}}
\affiliation{Universit\'{e} Paris-Saclay, CNRS/IN2P3, IJCLab, 91405 Orsay, France}
\author{J.~Billard}
\affiliation{Univ Lyon, Universit\'e Lyon 1, CNRS/IN2P3, IP2I-Lyon, F-69622, Villeurbanne, France}
\author{A.~Broniatowski}
\affiliation{Universit\'{e} Paris-Saclay, CNRS/IN2P3, IJCLab, 91405 Orsay, France}
\author{P.~Camus}
\affiliation{Institut N\'{e}el, CNRS/UJF, 25 rue des Martyrs, BP 166, 38042 Grenoble, France}
\author{A.~Cazes}
\affiliation{Univ Lyon, Universit\'e Lyon 1, CNRS/IN2P3, IP2I-Lyon, F-69622, Villeurbanne, France}
\author{M.~Chapellier}
\affiliation{Universit\'{e} Paris-Saclay, CNRS/IN2P3, IJCLab, 91405 Orsay, France}
\author{F.~Charlieux}
\affiliation{Univ Lyon, Universit\'e Lyon 1, CNRS/IN2P3, IP2I-Lyon, F-69622, Villeurbanne, France}
\author{M. De~J\'{e}sus}
\affiliation{Univ Lyon, Universit\'e Lyon 1, CNRS/IN2P3, IP2I-Lyon, F-69622, Villeurbanne, France}
\author{L.~Dumoulin}
\affiliation{Universit\'{e} Paris-Saclay, CNRS/IN2P3, IJCLab, 91405 Orsay, France}
\author{K.~Eitel}
\affiliation{Karlsruher Institut f\"{u}r Technologie, Institut f\"{u}r Astroteilchenphysik, Postfach 3640, 76021 Karlsruhe, Germany}
\author{J.-B.~Filippini}
\affiliation{Univ Lyon, Universit\'e Lyon 1, CNRS/IN2P3, IP2I-Lyon, F-69622, Villeurbanne, France}
\author{D.~Filosofov}
\affiliation{JINR, Laboratory of Nuclear Problems, Joliot-Curie 6, 141980 Dubna, Moscow Region, Russian Federation}
\author{J.~Gascon}
\affiliation{Univ Lyon, Universit\'e Lyon 1, CNRS/IN2P3, IP2I-Lyon, F-69622, Villeurbanne, France}
\author{A.~Giuliani}
\affiliation{Universit\'{e} Paris-Saclay, CNRS/IN2P3, IJCLab, 91405 Orsay, France}
\author{M.~Gros}
\affiliation{IRFU, CEA, Universit\'{e} Paris-Saclay, F-91191 Gif-sur-Yvette, France}
\author{E.~Guy}
\affiliation{Univ Lyon, Universit\'e Lyon 1, CNRS/IN2P3, IP2I-Lyon, F-69622, Villeurbanne, France}
\author{Y.~Jin}
\affiliation{C2N, CNRS, Univ.  Paris-Sud, Univ.  Paris-Saclay, 91120 Palaiseau, France}
\author{A.~Juillard}
\affiliation{Univ Lyon, Universit\'e Lyon 1, CNRS/IN2P3, IP2I-Lyon, F-69622, Villeurbanne, France}
\author{M.~Kleifges}
\affiliation{Karlsruher Institut f\"{u}r Technologie, Institut f\"{u}r Prozessdatenverarbeitung und Elektronik, Postfach 3640, 76021 Karlsruhe, Germany}
\author{H.~Lattaud}\email{lattaud@ipnl.in2p3.fr}
\affiliation{Univ Lyon, Universit\'e Lyon 1, CNRS/IN2P3, IP2I-Lyon, F-69622, Villeurbanne, France}
\author{S.~Marnieros}
\affiliation{Universit\'{e} Paris-Saclay, CNRS/IN2P3, IJCLab, 91405 Orsay, France}
\author{D.~Misiak}
\affiliation{Univ Lyon, Universit\'e Lyon 1, CNRS/IN2P3, IP2I-Lyon, F-69622, Villeurbanne, France}
\author{X.-F.~Navick}
\affiliation{IRFU, CEA, Universit\'{e} Paris-Saclay, F-91191 Gif-sur-Yvette, France}
\author{C.~Nones}
\affiliation{IRFU, CEA, Universit\'{e} Paris-Saclay, F-91191 Gif-sur-Yvette, France}
\author{E.~Olivieri}
\affiliation{Universit\'{e} Paris-Saclay, CNRS/IN2P3, IJCLab, 91405 Orsay, France}
\author{C.~Oriol}
\affiliation{Universit\'{e} Paris-Saclay, CNRS/IN2P3, IJCLab, 91405 Orsay, France}
\author{P.~Pari}
\affiliation{IRAMIS, CEA, Universit\'{e} Paris-Saclay, F-91191 Gif-sur-Yvette, France}
\author{B.~Paul}
\affiliation{IRFU, CEA, Universit\'{e} Paris-Saclay, F-91191 Gif-sur-Yvette, France}
\author{D.~Poda}
\affiliation{Universit\'{e} Paris-Saclay, CNRS/IN2P3, IJCLab, 91405 Orsay, France}
\author{S.~Rozov}
\affiliation{JINR, Laboratory of Nuclear Problems, Joliot-Curie 6, 141980 Dubna, Moscow Region, Russian Federation}
\author{T.~Salagnac}
\affiliation{Univ Lyon, Universit\'e Lyon 1, CNRS/IN2P3, IP2I-Lyon, F-69622, Villeurbanne, France}
\author{V.~Sanglard}
\affiliation{Univ Lyon, Universit\'e Lyon 1, CNRS/IN2P3, IP2I-Lyon, F-69622, Villeurbanne, France}
\author{L.~Vagneron}
\affiliation{Univ Lyon, Universit\'e Lyon 1, CNRS/IN2P3, IP2I-Lyon, F-69622, Villeurbanne, France}
\author{E.~Yakushev}
\affiliation{JINR, Laboratory of Nuclear Problems, Joliot-Curie 6, 141980 Dubna, Moscow Region, Russian Federation}
\author{A.~Zolotarova}
\affiliation{IRFU, CEA, Universit\'e Paris-Saclay, Saclay, F-91191 Gif-sur-Yvette, France}
\collaboration{EDELWEISS Collaboration} 

\author{B. J. Kavanagh}
\affiliation{Instituto de F\'isica de Cantabria (IFCA, UC-CSIC), Avenida de Los Castros s/n, 39005 Santander, Spain}

\date{\today} 

\begin{abstract}
The EDELWEISS collaboration reports on the search for Dark Matter (DM) particle interactions  via Migdal effect with masses between $32$~MeV$\cdot$c$^{-2}$ to $2$~GeV$\cdot$c$^{-2}$ using a $200$~g cryogenic Ge detector sensitive to simultaneously heat and ionization signals and operated underground at the  Laboratoire Souterrain de Modane in France. 
The phonon signal was read out using a Transition Edge Sensor made of a NbSi thin film.
The detector was biased at $66$~V in order to benefit from the  Neganov-Trofimov-Luke  amplification and resulting in a resolution on the energy of electron recoils of $4.46$~eV$_{ee}$ ($102.58$ eV at $66$~V ) (RMS) and an analysis threshold of  $30$~eV$_{ee}$. The sensitivity is limited by a dominant background not associated to charge creation in the detector. The search constrains a new region of parameter space for cross-sections down to $10^{-29}$~cm$^2$ and masses
between $32$ and $100$ ~MeV$\cdot$c$^{-2}$.
The achieved low threshold with the NbSi sensor shows the relevance of its use for out-of-equilibrium phonon sensitive devices for low-mass DM searches. 
\end{abstract}


\maketitle

\section{Introduction} \label{sec:intro}

The direct search for Dark Matter (DM) particle interactions with nuclei in a terrestrial \cite{Drukier:1984vhf,Goodman:1984dc,Drukier:1986tm} target has made huge progress for particles with masses in the range from 1 GeV$\cdot$c$^{-2}$ to $1$~TeV$\cdot$c$^{-2}$~\cite{xenon1t,lux,pandax}.
The absence of signal in that region has intensified the interest for the extension of the search to masses down to $1$~GeV$\cdot$c$^{-2}$ and below ~\cite{Essig,Cheung,Hooper,Falkowski,Petraki,Zurek,Bertone:2018krk}.
However, these lower masses raise additional experimental challenges: the need to lower the energy detection threshold well below 1~keV, the ionization or scintillation yield for nuclear recoil signals, and the appearance of new types of backgrounds at the lowest energies.
New detector designs targeting low energy threshold include cryogenic detectors~\cite{cresst,cdmslite,red30}, CCDs~\cite{damic,sensei} and low-threshold point-contact germanium ionization detectors~\cite{cedex}.
DM particles with masses below 1 GeV$\cdot$c$^{-2}$ and with large nucleon-elastic scattering cross-sections are yet to be excluded by direct searches, prompting some searches to be performed above ground~\cite{nucleus,red20,SuperCDMS:2020aus}. 

In parallel, it was noticed that the problem of the very low kinetic energy of the nuclear recoil, combined with the reduced ionization or scintillation yield for this type of event, could be circumvented by using the Migdal effect~\cite{Vergados:2004bm,Moustakidis:2005gx,Bernabei:2007jz,Ibe:2017yqa}.
This effect accounts for the probability that the collision between the DM particle and the nucleus may be accompanied by the release of an atomic electron. The energy imparted to the latter particle is typically much larger than the kinetic energy of the nuclear recoil~\cite{Ibe:2017yqa}, and thus easier to detect. 
In addition, the ionization yield of the electron is not affected by quenching, resulting in important improvements in the sensitivity of ionization-based searches for DM particles with masses below $200$~MeV$\cdot$c$^{-2}$~\cite{Dolan:2017xbu,Akerib:2018hck,xenon-migdal,cedex}. 
Although the Midgal effect has never been directly observed in a nuclear collision, and  precise calculations require special care for electrons in the outermost electronic shells in solids~\cite{outer-essig,outer-liu},  upper limits on DM-nucleus collision rates can be extracted from the calculation involving electrons below the valence shell.

Based on the Midgal effect, the EDELWEISS collaboration was able to extend the mass range for the search of DM particles down to $45$~MeV$\cdot$c$^{-2}$, using a $33.4$~g cryogenic Ge detector equipped with a Ge Neutron Transmutation Doped (Ge-NTD) thermal sensor~\cite{red20}.
That range was limited by the energy threshold of 60 eV. The excluded cross-sections were constrained by the background, originating from the poorly shielded, above-ground environment of the detector, but also by a large population of events. Later studies~\cite{red30} suggested that this population  is not associated with the creation of electron-hole pairs, and has been so-called Heat-Only (HO) events.
In this paper, we present the results of DM searches using  Midgal effect performed in a low-radioactivity underground environment with a cryogenic detector equipped with a NbSi Transition Edge Sensor (TES), developed to reduce the sensitivity of EDELWEISS detectors to HO events, and with a low energy threshold for electron recoil.

The paper is organized as follows. In Sec. II, we present the detector and experimental setup. In Sec. III, we give details of the DM search, including data processing, detector calibration and data analysis. In Sec. IV, we present the extracted limits on DM particles using the Migdal effect. Finally, we conclude in Sec. V.
\section{Experimental setup} \label{sec:edelweiss}

The DM search was performed at the \textit{Laboratoire Souterrain de Modane} (LSM, France) using the ultra-low background environment of the EDELWEISS-III cryostat~\cite{edwtech}. 
The detectors are thus protected by a 4800 m.w.e. rock overburden, an outer polyethylene shield  of 50 cm, followed by a 20 cm lead shield with an inner layer of $2$~cm of roman lead and an inner polyethylene shield with an average thickness of 10 cm.

\begin{figure}[!h]
\begin{center}
\includegraphics[width=0.99\linewidth]{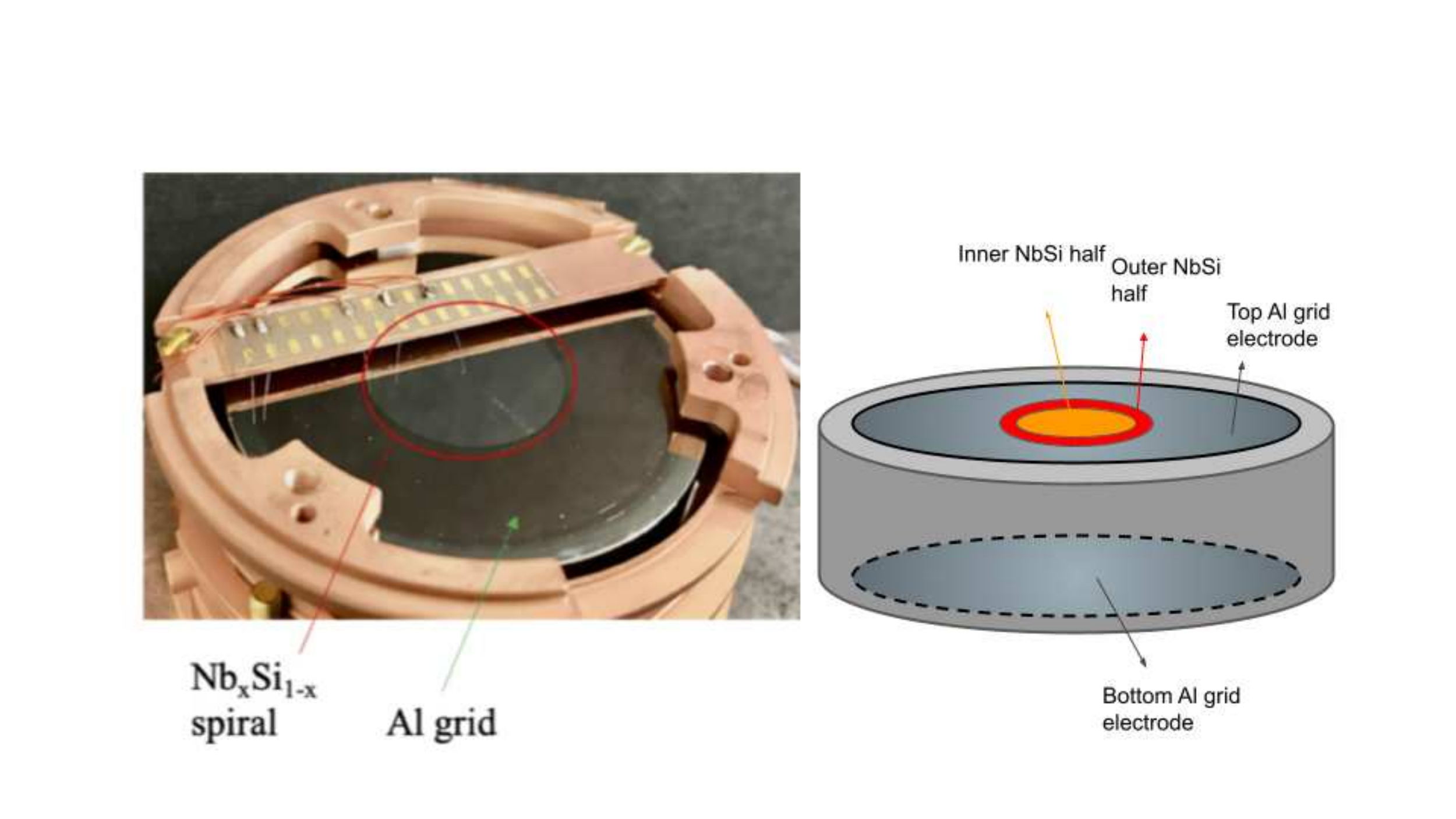}
\caption{Left: Top side of the NbSi209 200~g detector equipped with a NbSi TES, inside its copper holder. The red circle indicates the position of the 20 mm diameter NbSi sensor. The outer diameter of the crystal is covered with the Al mesh electrode. Right: sketch of NbSi209 detector with the outer and inner  halves of equal resistance of the NbSi sensor in red and orange respectively, the Ge crystal in light gray and the Al electrode in dark gray. }\label{nbsi_picture}
\end{center}
\end{figure}

In an attempt to better understand and consequently control the important background of HO events that affects previous EDELWEISS detectors equipped with Ge-NTD thermistors, a new type of sensor of different design and concept was used in the present search. Beyond material differences between the two sensors, they differ in their sensitivity to thermal and out-of-equilibrium phonons, and could reveal differences in the mechanisms in the formation of the HO and normal signals.
The detector used for the search named NbSi209 is a $200$~g Ge cylindrical crystal (48~mm in diameter and $20$~mm in height) on top of which was lithographed a Nb$_x$Si$_{1-x}$ thin film TES~\cite{nbsi-ltd}. The 100~nm-thick film is shaped as a spiral with a track width of 160 $\mu$m. Fig.~\ref{nbsi_picture} shows the top side of the detector. 
The film is maintained near the temperature of 44 mK, at which its transition between the superconducting and normal state occurs, with the help of a heater resistance hanged to the copper holder and linked  to the detector through gold wire. 
In its normal state, the film resistance is 2~M$\Omega$.
The spiral is split in two equal-resistance halves, resulting in a central phonon sensor with a diameter of 14~mm and an annular ring sensor of 3~mm in width (see Fig.~\ref{nbsi_picture} left), those two halves are read as independent channel.
The voltage injected across each TES half is continuously read out using the standard EDELWEISS-III cold-FET based electronics at 100~K~\cite{edwtech}. The TES are read with a square current with an intensity of the order of $1$ nA modulated at $500$ Hz. This current induced a small bias of the order of $0.1$ mV. 
In contrast with the Ge-NTD sensor used in previous EDELWEISS detectors~\cite{edwtech,red20,red30}, the TES technology has been shown to be sensitive to out-of-equilibrium phonons~\cite{nbsi-ltd}.
The signal has a rise time of less than 1~ms, and two decay constants of 10 and 100~ms, associated to the electron-phonon time constant in the film and the thermal leak of the detector, respectively.

The heat link between the detector and the copper holder is done via gold wires bonded on a gold pad located at the center of the bottom flat surface of the detector.
To preserve the lifetime of out-of-equilibrium phonons inside the Ge absorber, the electrodes covering the two flat surfaces are $20$~nm thick Al grids with a $4\%$ coverage fraction 
($10~\mu$m lines with a $500$~nm pitch).
The top electrode covers the outer annulus beyond the NbSi film, and it is held at a bias of 0~V. 
The bottom side is fully covered by an electrode biased at a voltage varying between $\pm 66$~V.
The electrodes are read out separately. 
The charge collected on the NbSi film is not read out. 
In addition to collecting charge, the bias applied to the electrodes is such that the drift of $N$ electron-hole pairs across a voltage difference $\Delta V $ produces additional phonons. Those phonon energies $E_{NTL}$ =  $Ne\Delta V$ ($e$ is the elementary charge) add to the initial recoil energy. This is the  so-called Neganov-Trofimov-Luke (NTL) effect\cite{Neganov,Luke}. It essentially turns a cryogenic calorimeter (operated at $\Delta V $ = 0 V) into a charge amplifier with a mean gain $\langle g \rangle = (1+e\Delta V /\epsilon_{gamma})$, where $\epsilon_{gamma} = 3.0$~eV is the average ionization energy in Ge for electron recoils~\cite{knoll}.

The data acquisition system and readout electronics are the same as in~\cite{edwtech}. The data from the phonon and ionization channels were digitized at a frequency of $100$~kHz, filtered, and continuously stored on disk with a digitization rate of $500$~Hz.
Data were collected between December 2018 and July 2020, during the same cool-down as in ~\cite{red30}.
The beginning of that time period was devoted to optimize the film working temperature that results in the maximum signal-to-noise ratio when the two TES signals are combined linearly. 
These conditions are found optimal at a temperature of $44$~mK, for inner and outer TES resistance values of 100 and $500$~k$\Omega$, respectively.
The corresponding resolutions for each sensor are approximately 130 eV, resulting in a combined resolution between $90$ and $100$~eV, and combination factors close to $50\%$.
These values do not vary significantly with the bias value, within the $\pm66$~V range fixed by the electronics~\cite{edwtech}.
For the NTL gain of $\langle g \rangle=23$ obtained for electronic recoil with a bias of $66$~V, this corresponds to a resolution of approximately $4$~eV electron-equivalent (eV$_{ee})$.
The accumulation of trapped charges in the detector is controlled by applying the same method as in Ref.~\cite{red30}.

\begin{figure}[!h]
\begin{center}
\includegraphics[width=\linewidth]{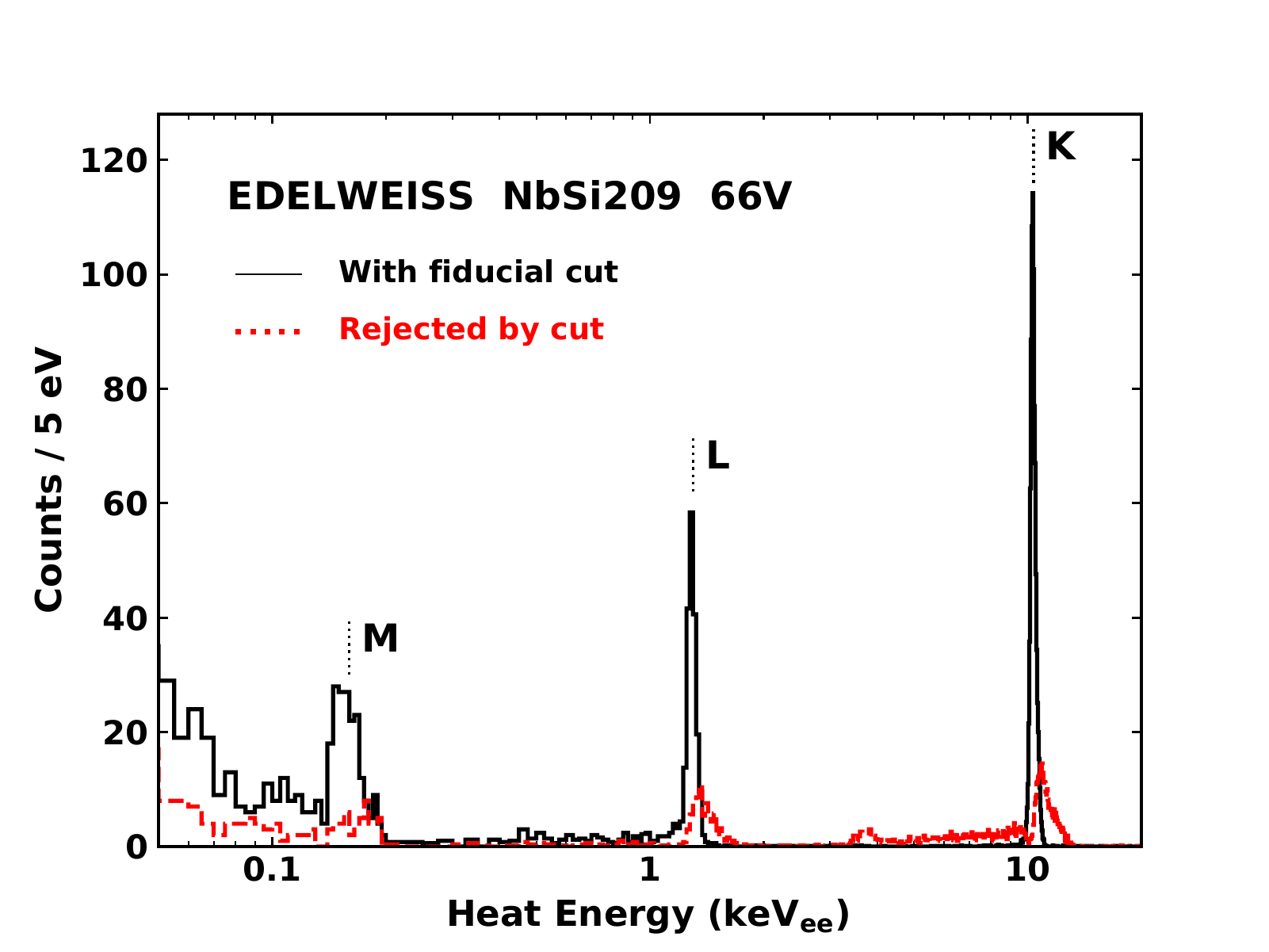}
\caption{ Energy spectrum recorded at a bias of $66$~V following  the $^{71}$Ge emitted x-ray lines induced by the neutron activation of the Ge detector, resulting in the strong lines characteristic of $^{71}$Ge x-ray emissions. Black: events selected with the fiducial selection discussed in the text. Red: events rejected by the selection.}\label{KLM_graph}
\end{center}
\end{figure}

In April 2019, the detector was uniformly activated using a strong AmBe neutron  source. The produced short-lived $^{71}$Ge isotope  decays by electron capture in the K, L, and M shells, with de-excitation x-ray lines at $10.37$, $1.30$, and $0.16$~keV, respectively. The low energy x-ray lines are locally absorbed, thus providing very good probes of the detector response to a DM signal uniformly distributed inside the detector volume. 
These are clearly visible in Fig.~\ref{KLM_graph}, which shows the energy spectrum for the phonon signal from calibration data recorded at a bias of $66$~V. 

On that figure, the solid black histogram represents events where the charges collected on both electrodes are equal within the ionization measurement precision ($\sigma = 200$~eV$_{ee}$), and the dashed red one, those where the two charge collections differ.
This corresponds to two populations. The first population are events occurring in the sub-cylinder volume facing the area delimited by the imprint of the NbSi film, easily tagged by the fact that the bottom electrode collects the entirety of the downward moving charges but the upward-moving ones end their drift in the NbSi film and not including a signal on the top electrode.
The second one consists of events where a significant fraction of the charges end up on the cylindrical edge of the detector, and the trapped ones produce an asymmetric signal on the two electrodes. With these tags based on the ionization signals, we observe that the first population corresponds to the tail at the right of the 10.37 keV peak in Fig.~\ref{KLM_graph}, while the second one corresponds to the tail in between the two peaks.
The same pattern is also observed for the $1.3$~keV$_{ee}$ peak.
The black Gaussian peak at $10.37$~keV$_{ee}$ thus corresponds to events occurring inside the volume defined by field lines leading to the top electrode and sufficiently far away from the outer edge of the detector, representing 63\% of all K-shell events. 

Between $30$ and $200$~eV$_{ee}$, the energy range relevant for the Migdal DM search, the ionization criterion (same charge collection on both electrode) cannot reliably separate these populations.
Therefore, the criterion will not be applied. 
In Sec.~\ref{sec:DMsearch}, the efficiency associated with events in the $10.37$-keV peak  will be kept as a conservative lower limit on the total efficiency of the detector.

\section{Dark matter Search} \label{sec:DMsearch}
The DM search has been performed using the data set recorded when the detector was operated at $66$~V and selecting only time periods when the phonon baseline resolution is less than  $140$~eV RMS. This dataset was recorded from March 2019 to June 2020. The average baseline heat energy resolution in the search sample is 102 $\pm$ 12~eV RMS, corresponding to  4.46 $\pm$ 0.54~eV$_{ee}$ RMS once the NTL gain $\langle g \rangle$ is considered. For the ionization channel, the resolution is 210.3 $\pm$ 16.3~eV RMS, and stable over time.


The $^{71}$Ge peaks have been used for the calibration of the heat and ionization samples in the four months following the AmBe activation. The ionization calibration was observed to be constant over that period. Calibrations with the $356$~keV gamma-ray of a $^{133}$Ba source at the beginning, middle and end of the 19-month data-taking period confirmed this stability. The gain of the heat signal was observed to vary slowly by $\pm 10$\% depending on cryogenic conditions. This was corrected with a precision better than $1$\% by monitoring the ratio of the heat and ionization signals of events between 5 and 60 keV, and with a precision of $0.1$\% in samples where the $^{71}$Ge peaks are observed. The non-linearity of the heat channel was measured using the position of the $^{71}$Ge KLM activation peaks observed at different NTL amplification. It is $5$\% between $1$ and $500$~keV, and the precision of the correction at low energy is better than $2$\%

In order to set the analysis selection criteria, one out of every two hours of data were blinded  and  the other half kept  to set the analysis selection criteria and excluded from the search. In order to derive conservative constraints on DM interaction, it was decided not to subtract possible backgrounds. 
This represents $27.9$ days of blinded data and $28.8$ days of non-blinded data for the reference sample.

The offline trigger is based on an optimal matching filter approach, which is essentially the same procedure as described in \cite{red20}. The numerical procedure used is detailed in \cite{trigger}. The pulses are searched iteratively in the filtered data stream using a decreasing energy ordering criterion. 
This will induce an energy dependency in the triggering efficiency, especially for low energy events for which the dead time is larger than for high amplitude pulses.
In this algorithm, a time window of $\Delta t= 2.048$~s is allocated around the pulse with the largest amplitude. This time window is then excluded from the process at the next iteration, it continues until there is no time interval larger than   $\Delta t$ in the data stream.
In order to assess the amplitude of the pulses, they are fitted in the frequency domain by minimizing a $\chi ^2  $ function, based on the standard pulse shape derived from K-peak events.

\begin{figure}[!h]
\begin{center}
\includegraphics[width=0.99\linewidth]{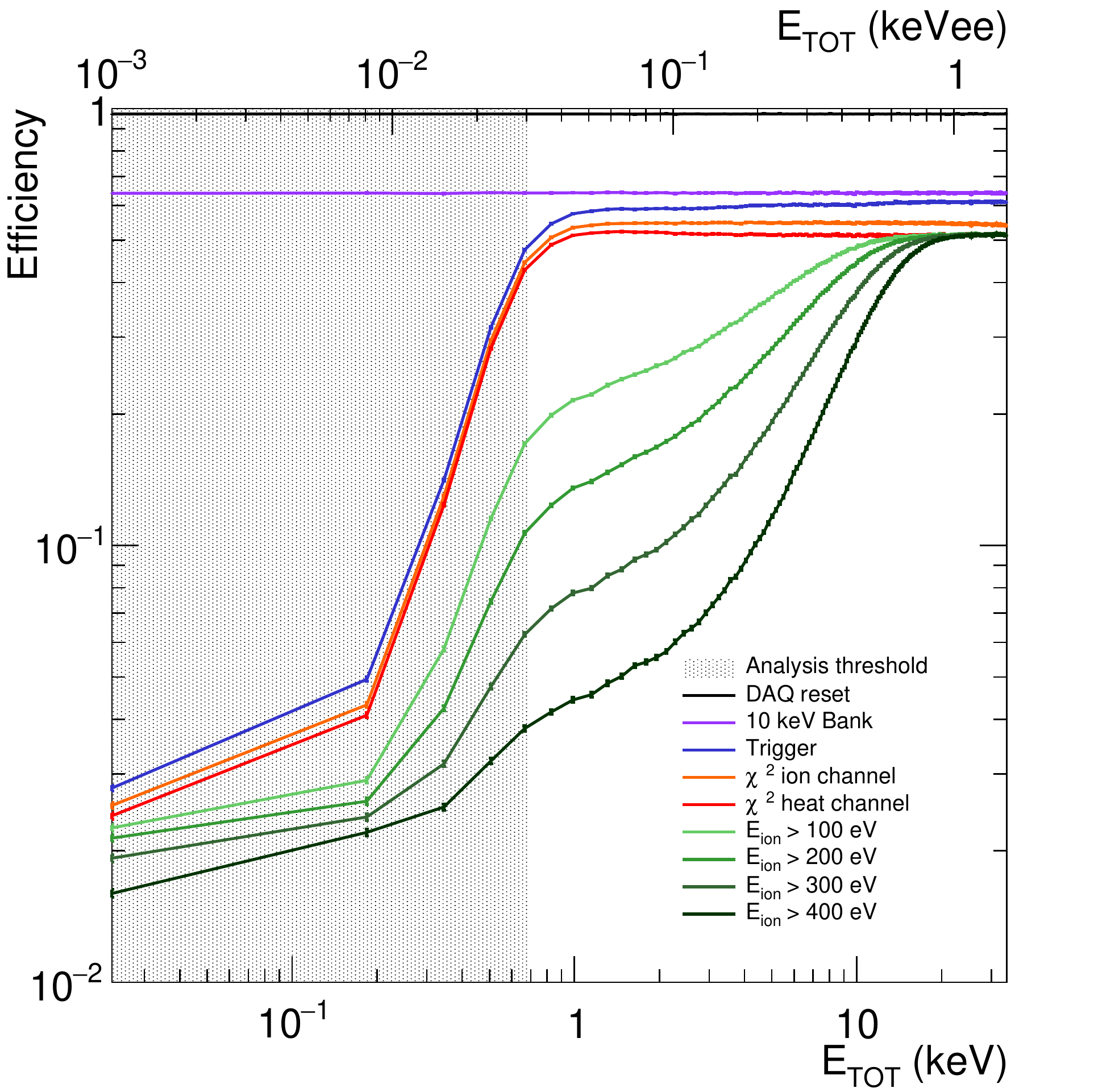}
\caption{Signal efficiency (fraction of events surviving selection criteria) as a function of the input phonon heat energy of events inserted in the data stream in units of keV (lower axis), or keV-electron-equivalent (keV$_{ee}$ (upper axis)). The different color lines correspond to different steps in the trigger and data analysis selection, the analysis threshold is shown in gray (see text). The uncertainties associated with each point correspond to binomial uncertainties.}\label{efficiency_graph}
\end{center}
\end{figure}

The dedicated data-driven method used to carefully model this effect and estimate the efficiency energy dependency of the analysis is described in detail in~\cite{red20} and summarized below.
Pulses of known energy randomly selected among  the events from the K-line decay are scaled to relevant energies and injected in the data stream at random times. 
This bank of events contains $10667$ traces of K-shell events with  energies between $2$ and $12.6$~keV$_{ee}$; recorded at $66$~V after the activation of the detector. This bank contains all types of events occurring in the detector, including events with incomplete charge collection and events with extra energy from out-of-equilibrium phonons. 

Those real pulses are first scaled down to a fraction of  $10.37$~keV$_{ee}$, in order to estimate efficiency at the desired energy,  and injected at a rate of $ 0.02$~Hz to minimize  the induced dead time below $1 \%$ of the total trigger dead time rate. The ionization pulses are scaled by the same factor and included in the procedure to take into account the biases induced by the pulse fitting procedure, which yield to a systematic shift of the ionization resolution from  $210$ to $225$ eV.

Fig.~\ref{efficiency_graph} shows the efficiency as a function of the scaled and injected pulse energies at various stages of the triggering and data selection procedure. The first criterion (black line) corresponds to the DAQ electronic resets that induce a dead time  of $\sim 2.8\%$.  
Since the goal is to provide upper limits on event rates,  it was decided to only take into account  the contribution to the total efficiency  of the $65\%$ of event present in the K-peak (events with energies  $\in [9.6,11]$~keVee), the associated $35\%$ drop in efficiency across all the energy range is shown by the purple line. This gives a lower limit on the efficiency that can underestimate the signal up to $50\%$ as it effectively treats part of the potential signal as any other background, but it has the advantage of not depending on the detailed understanding and modeling of the tails of the K-peak due to incomplete charge collection and additional energy from out-of-equilibrium phonons.
The efficiency of the trigger procedure as a function of the injected energy is described by the blue line. 
The slightly rising curve slope between $0.8$ and $10$~keV in Fig.~\ref{efficiency_graph} is due to the bias of the trigger algorithm favoring high energy events, increasing the dead time for low energy events. The much steeper curve slope between $0.02$ and $0.8$~keV reflects the large number of events coming from electronic noise. 
The orange and red lines in Fig.~\ref{efficiency_graph} correspond to the criterion applied on the pulse shape of the ionization and heat signals respectively, through the fitted $\chi^{2}$ which help to reduce contamination from pileup and badly reconstructed events. 
With these criteria, the plateau efficiency is obtained at $1$~keV ($30$~eV$_{ee}$).
This threshold is considerably better than the one achievable with the ionization channel alone. However, the following analysis  considers different criteria on the ionization energy   whose impact is shown in Fig.~\ref{efficiency_graph}.

As can be seen in  Fig.~\ref{KLM_graph}, there is a rise in the event rate below $200$~eV$_{ee}$ ($4.6$~keV). As will be shown in the discussion (see Sec.~\ref{sec:discussion}), these events are not associated with the production of electron-hole pairs in the detector, and are hence so-called as Heat-Only events. Those events are pure heat as no ionization is associated with their heat energy deposit. 
This population can be reduced by requiring the presence of a significant ionization signal. However, the performance of such criterion at low energy is limited by the ionization resolution of $210$~eV.
The green lines in Fig.~\ref{efficiency_graph} show the effect of requiring a minimal ionization signal of $100$, $200$, $300$ or $400$~eV on the efficiency.
The efficiency loss corresponds to what is expected by the observed Gaussian noise of ionization signals.
The optimization of this criterion will be discussed in Sec.~\ref{sec:results}.

\begin{figure}[!h]
\begin{center}
\includegraphics[width=0.99\linewidth]{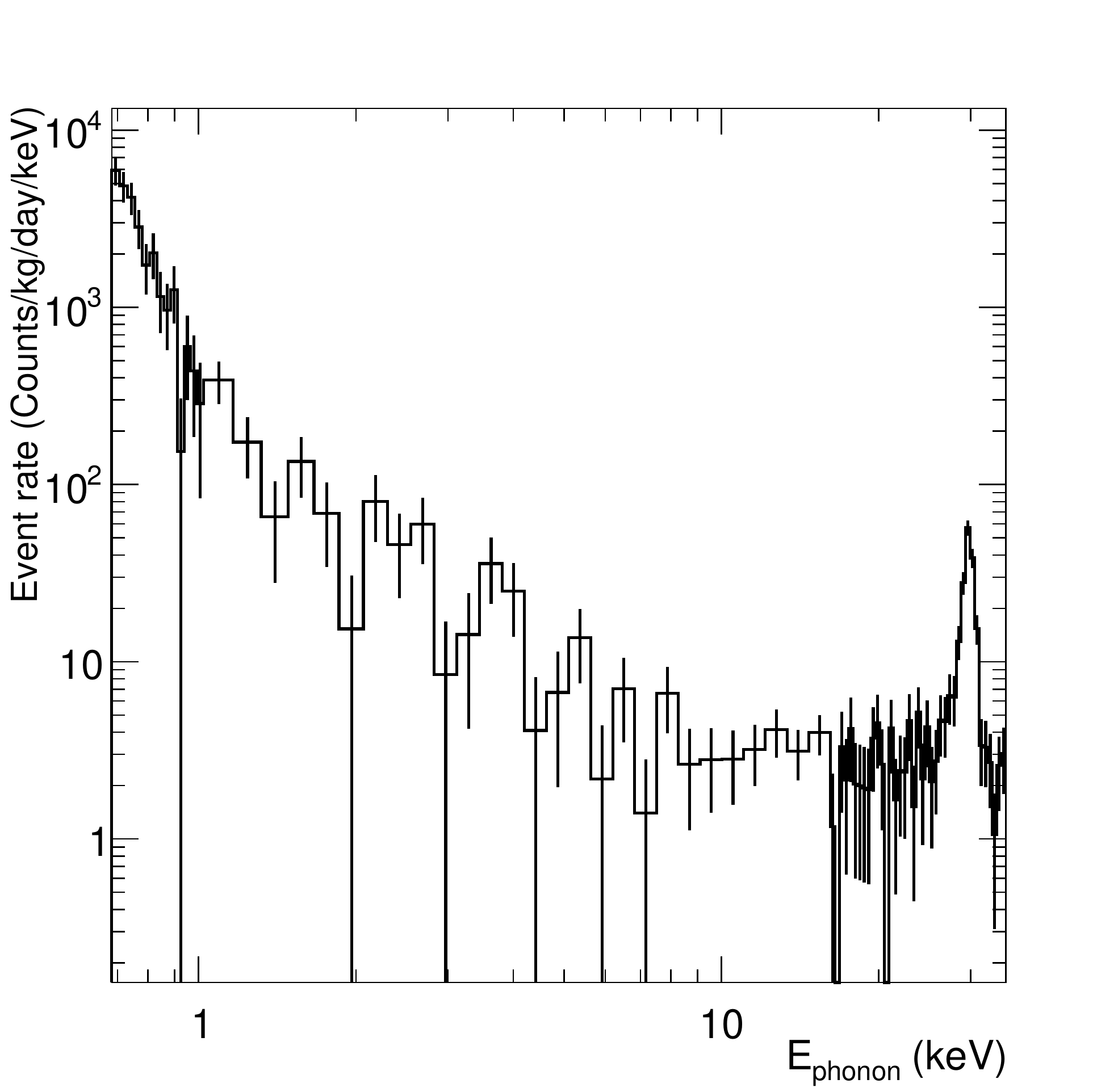}
\caption{Efficiency-corrected energy spectrum of the blinded part of the dataset after nominal analysis data selection and $E_{ion}>400$~eV criterion (corresponding to the efficiency curve in dark green in Fig~\ref{efficiency_graph}). }\label{FinalspectrumDRU}
\end{center}
\end{figure}

The  efficiency-corrected spectrum of the blinded dataset is shown in Fig.~\ref{FinalspectrumDRU}, the efficiency curve applied is the one corresponding to the criterion requiring more than $400$~eV of ionization energy  (darkest green curve in Fig~\ref{efficiency_graph}). This efficiency curve will be the one applied to the signal in Sec~\ref{sec:results}. At high energy, the spectrum is dominated by the $1.3$~keV$_{ee}$ L-peak shifted up to 29.9 keV after the NTL boost.

\section{Results} \label{sec:results}

The data shown in Fig.~\ref{FinalspectrumDRU} have been interpreted in terms of  limits on the spin-independent interaction of DM particles with target atoms through the so-called Migdal effect. 
This effect states that an interaction of a DM particle with an atom may induce simultaneously a nuclear recoil and the ionization of an electron.
Calculations for non-isolated atoms in semiconductors have been performed \cite{2021semicondmigdal,Essig:2019xkx,Hochberg:2021pkt}. They show that the Migdal effect at low energy is enhanced in semiconductor with respect to isolated atom. These calculations are still evolving. However, in Ge, the contribution to this effect comes from $n=4$ valence shell electrons and mostly yields signals below $30$ eV. 
Consequently, as done in \cite{red20}, we use instead the widely used calculations from \cite{Ibe:2017yqa,outer-essig}. Those isolated-atom calculations do not take into account the  full band structure of Ge in a crystal, and therefore only the contribution of the $n=3$ shell is considered, in order to yield conservative rate limits. As this shell gives no signal below $30$ eV$_{ee}$, an analysis threshold is set at this energy, corresponding to a phonon energy of $690$ eV.
The contribution from $n\leq2$ shells  has also been neglected, since it does not yield an exploitable signal in our detector because of the very low probability of emitting an electron from this shell. 

As in \cite{red20}, the spin-independent DM-nucleus interactions are described using the standard astrophysical parameters for a Maxwellian velocity distribution \cite{standardassumptions} with an asymptotic velocity $v_0=220$ km$\cdot$s$^{-1}$ and a galactic escape velocity $v_{esc} = 544$~km$\cdot$s$^{-1}$, combined with a lab velocity $v_{lab}=232$~km$\cdot$s$^{-1}$. The local DM density is assumed to be $\rho _0 = 0.3$~GeV$\cdot$c$^{-2}\cdot$cm$^{-3}$.
The loss in coherence at high momentum is taken into account with the standard Helm form factor \cite{Helm:1956zz}. It is assumed that the cross-section scales as $A^2$, with $A$ being the mass number of the considered nucleus \cite{Digman:2019wdm}.
For a $100$ MeV$\cdot$c$^{-2}$ WIMP, an induced nuclear recoil has less than $1$~eV in energy, a quantity further reduced by some quenching factor, typically $0.1$ at these low energies \cite{quenchingpaper}, whereas a Migdal electron  yields a $~100$ ~eV signal. In order to avoid systematic uncertainties linked to the quenching factor, which is not properly measured for such low energy nuclear recoils~\cite{quenchingpaper}, only the electronic contributions to the signal energy are considered in the following.

Because of the experiment underground location, the DM energy spectrum and flux will be modified by the action of the stopping power of the rock overburden \cite{Kavanagh:2016pyr, Emken:2017qmp, Mahdawi:2018euy, Hooper:2018bfw}.
These Earth-shielding effects were calculated using the publicly available \textbf{verne} code \cite{verne}, introduced in \cite{Kavanagh:2017cru}. A continuous energy loss of the DM particles is assumed through the atmosphere, the 1700 m rock overburden and  the $20$~cm lead shielding, as well as straight line trajectories \cite{Starkman:1990nj}. Ref.~\cite{Emken:2018run} has shown that this simplified formalism gives rise to constraints similar to more complete Monte Carlo simulations. 

For DM particles moving at low velocities, near the escape velocity at the Earth surface, $v_{earth} = 11$~km$\cdot$s$^{-1}$, effects such as gravitational capture \cite{Gould:1987ww,Mack:2007xj} and gravitational focusing \cite{Kouvaris:2015xga} are not negligible. These effects are not taken into account in the flux calculation. Instead, the DM velocity distribution is conservatively set to zero below $v_{cut} =20$~km$\cdot$s$^{-1}$ when calculating the signal spectra. 
As in \cite{red30}, the detector response to these calculated signals is simulated using the pulse simulation procedure presented in Sec.~\ref{sec:DMsearch}.

\begin{figure}[!h]
\begin{center}
\includegraphics[width=0.99\linewidth]{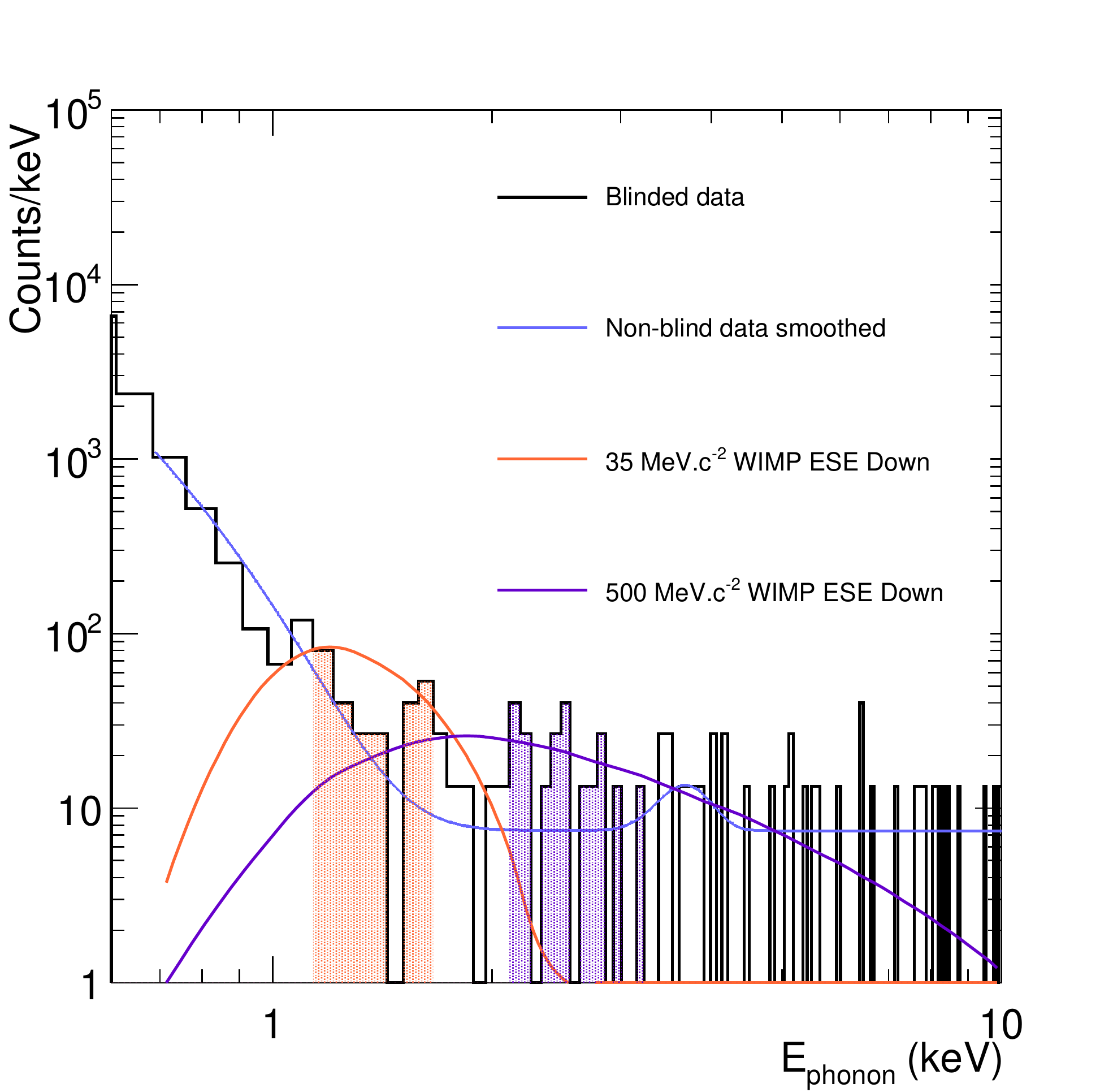}
\caption{Energy spectrum after selection  for the blinded data set in counts per keV (black histogram). Reference sample data smoothed by analytical function (plain blue). The curves show the  excluded  Migdal spectra smeared to detector resolution, corrected for the Earth shielding effect (ESE) and efficiency corrected for  WIMPs of $35$ (orange) and $500$ (purple) MeV$\cdot$c$^{-2}$ and the associated RoIs. }\label{figspectrum}
\end{center}

\end{figure}
The signal occurs in a region of the spectrum where reliable background models are not available, as this is the first time that it is explored with the present detector technology. Therefore, the physics reach of this first prototype is estimated with a search limited to establishing the signal rate that is excluded at $90\%$~C.L.  by the observed spectrum using Poisson statistics, without any background subtraction. 
For each DM mass, the signal rate is calculated within a region of interest  (RoI). This RoI is selected to maximize the sensitivity to the calculated signal of a hypothetical experiment, where the expected spectrum is taken as a smoothed version of the reference sample.

To reduce the effect of statistical fluctuations in the reference sample on the determination of the RoI, its energy spectrum has been smoothed using a sum of exponential functions together with a Gaussian peak to account for the presence of the $160$~eV line from $^{71}$Ge neutron activation.  A minimum width of $23$~eV is imposed to the RoIs. 
The optimization of the RoIs is repeated for different values of the  ionization energy criterion presented in Sec.~\ref{sec:DMsearch}, in order to achieve the best sensitivities. The optimized value for the  criterion which improves up to a factor $3$ the achieved limits is $E_{ion} > 400$~eV, shown as the darkest green in Fig.~\ref{efficiency_graph}. With a large but acceptable statistical cost, the cut reduces the efficiency corrected rate by a factor  $2$  at $1$~keV.
Once RoIs are fixed for each DM-mass using the reference sample, the $90 \%$~C.L. upper limit on a possible Migdal signal is calculated using Poisson statistics, again considering that all events in the search data sample RoIs are potential DM candidates. 

\begin{figure}[!h]
\begin{center}
\includegraphics[width=0.99\linewidth]{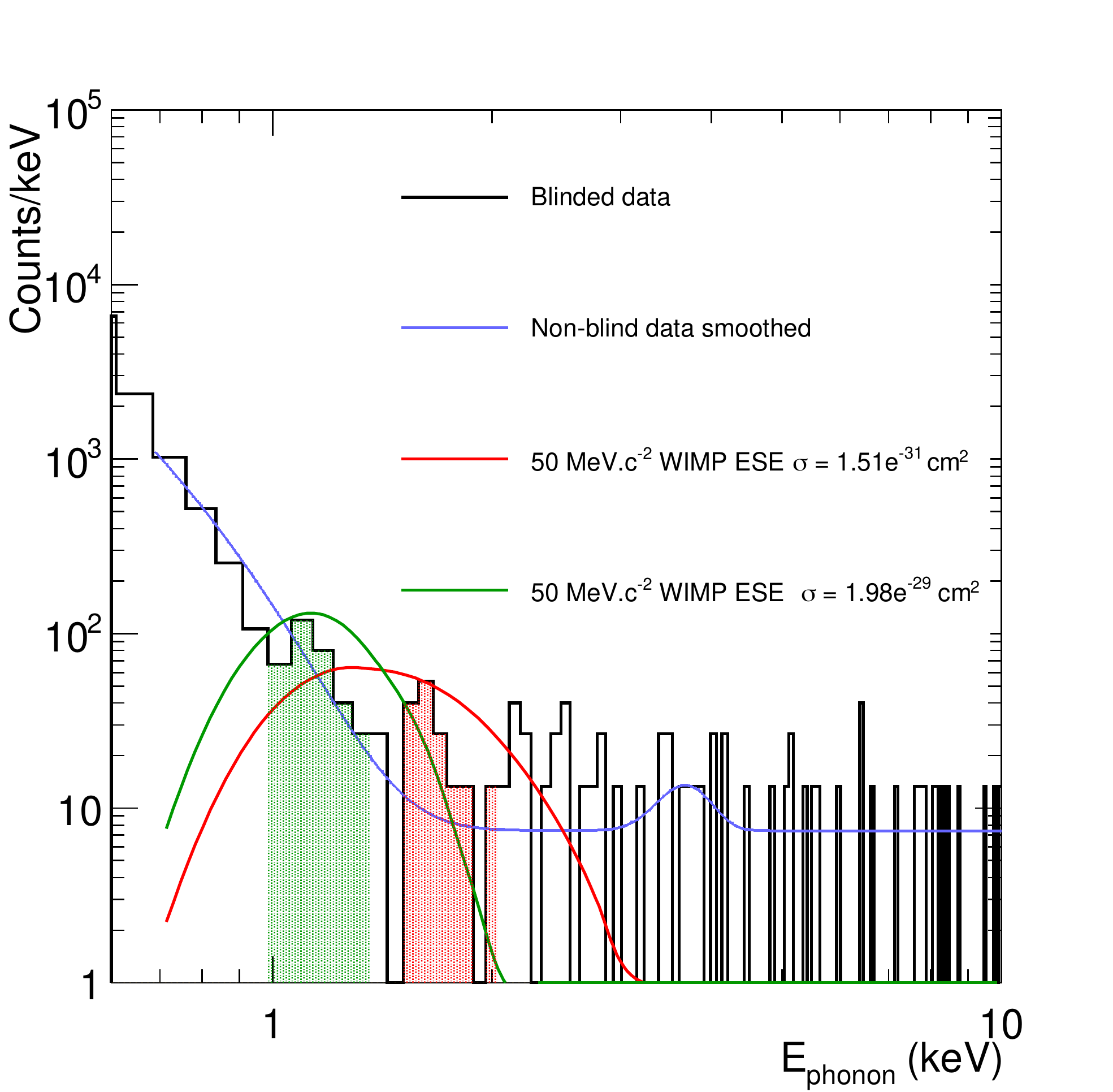}
\caption{Energy spectrum after all criteria applied for the blinded data set in counts per keV (black histogram). Reference sample data smoothed by analytical function (plain blue). The red and green curves show the  excluded  Migdal spectra smeared to detector resolution, corrected for the Earth-shielding effect and efficiency corrected  for  WIMPs of $50$~MeV$\cdot$c$^{-2}$ for lower and upper excluded cross section, respectively.  }\label{limit_graphESE}
\end{center}

\end{figure}

The resulting distributions for DM masses of 35 and $500$~MeV$\cdot$c$^{-2}$ and the associated RoIs are shown in Fig.~\ref{figspectrum}, where they are compared to the same experimental data as in Fig.~\ref{FinalspectrumDRU} (presented in this case without the efficiency correction and in count per keV) on which is overlaid  a smoothed model extracted from the independent reference sample (in blue). The model thus includes a Gaussian peak at $3.7$~keV ($160$ eV$_{ee}$) corresponding to the Ge M line.
This shows how the signal drifts towards high energies for higher DM particle masses.
The $90\%$~C.L. limits are calculated for both the lowest excluded cross-section, but also for the highest cross-section for which Earth-Shielding would prevent the observation of a signal in the detector. 
These two types of excluded signals for a $50$~MeV$\cdot$c$^{-2}$ WIMP are shown with their associated RoIs in Fig.~\ref{limit_graphESE}, in green for the upper part of the contour and in red for the lower one. 

\begin{figure}[!h]
\begin{center}
\includegraphics[width=0.99\linewidth]{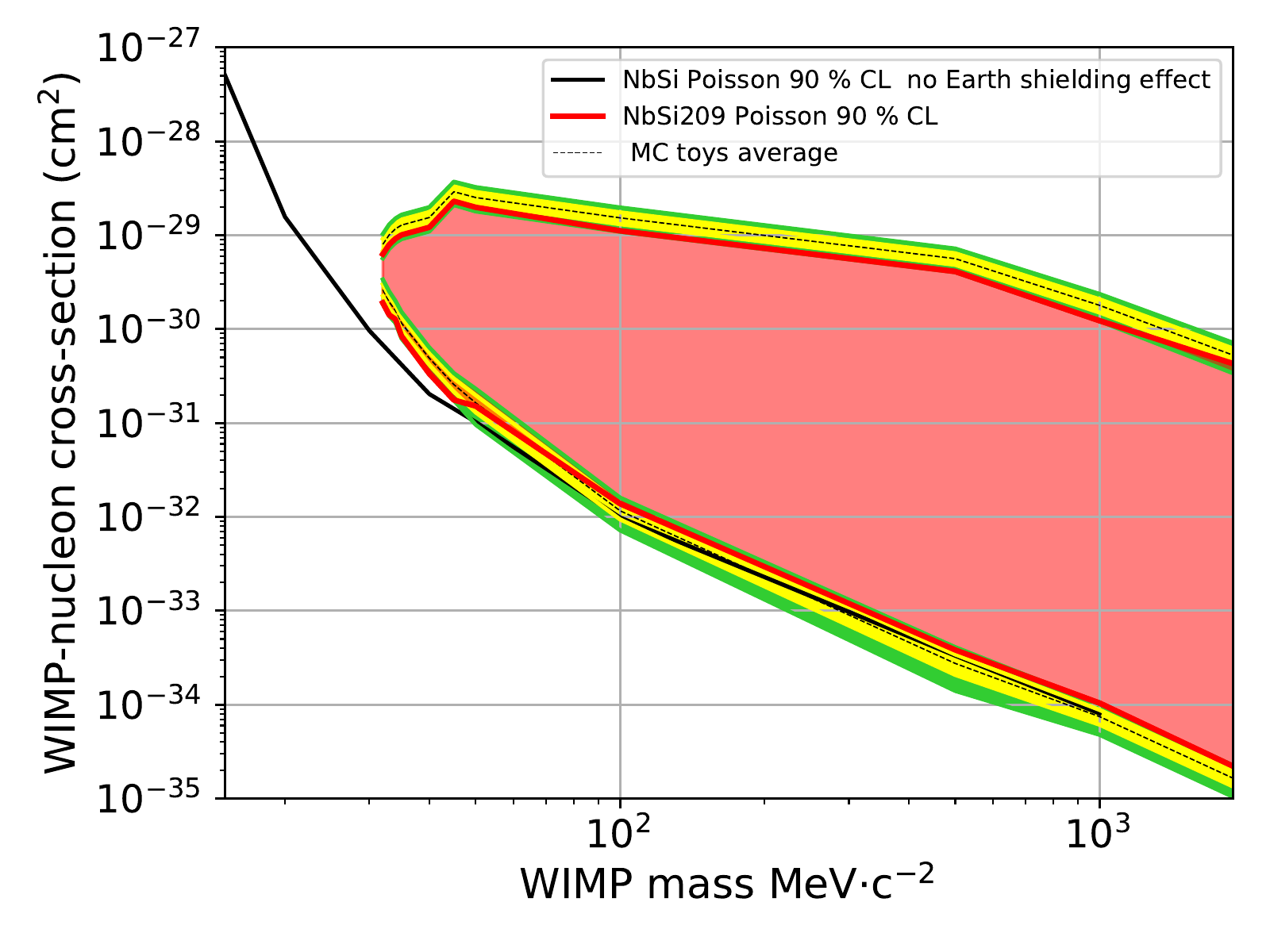}
\caption{$90\%$ C.L. upper limit on the cross-section for Spin-Independent interaction  between DM and Ge nuclei through Migdal effect. The black curve does not account for Earth shielding effect, the red contour is obtained by taking into account the slowing of the DM particle flux through the material above the detector. The yellow (green) envelope corresponds to the $1 \sigma$ ($2 \sigma$) statistical fluctuation estimated with Monte Carlo toys based on data model.}\label{figeesevsstd}
\end{center}

\end{figure}

The extracted limits are shown in Fig.~\ref{figeesevsstd}.
The red contour (delimited by the thick red line) corresponds to the excluded cross-sections for WIMP masses from $32$~MeV$\cdot$c$^{-2}$ to $2$~GeV$\cdot$c$^{-2}$. The yellow and green bands correspond to the statistical uncertainties at $1$ and $2\sigma$ determined using a Monte Carlo simulation randomly drawing events from the distributions of the signal and the smoothed reference. 
This shows that the red contour behaves as an expected fluctuation from our procedure
with respect to the statistics of the search sample. The black line shows the $90 \%$~C.L. upper limit on the Migdal DM interaction for signal neglecting the effect from Earth shielding.
The comparison with the red contour shows that these effects modify the rate and shape of the spectra for masses lower than $50$~MeV$\cdot$c$^{-2}$.

\begin{figure}[!h]
\begin{center}
\includegraphics[width=0.99\linewidth]{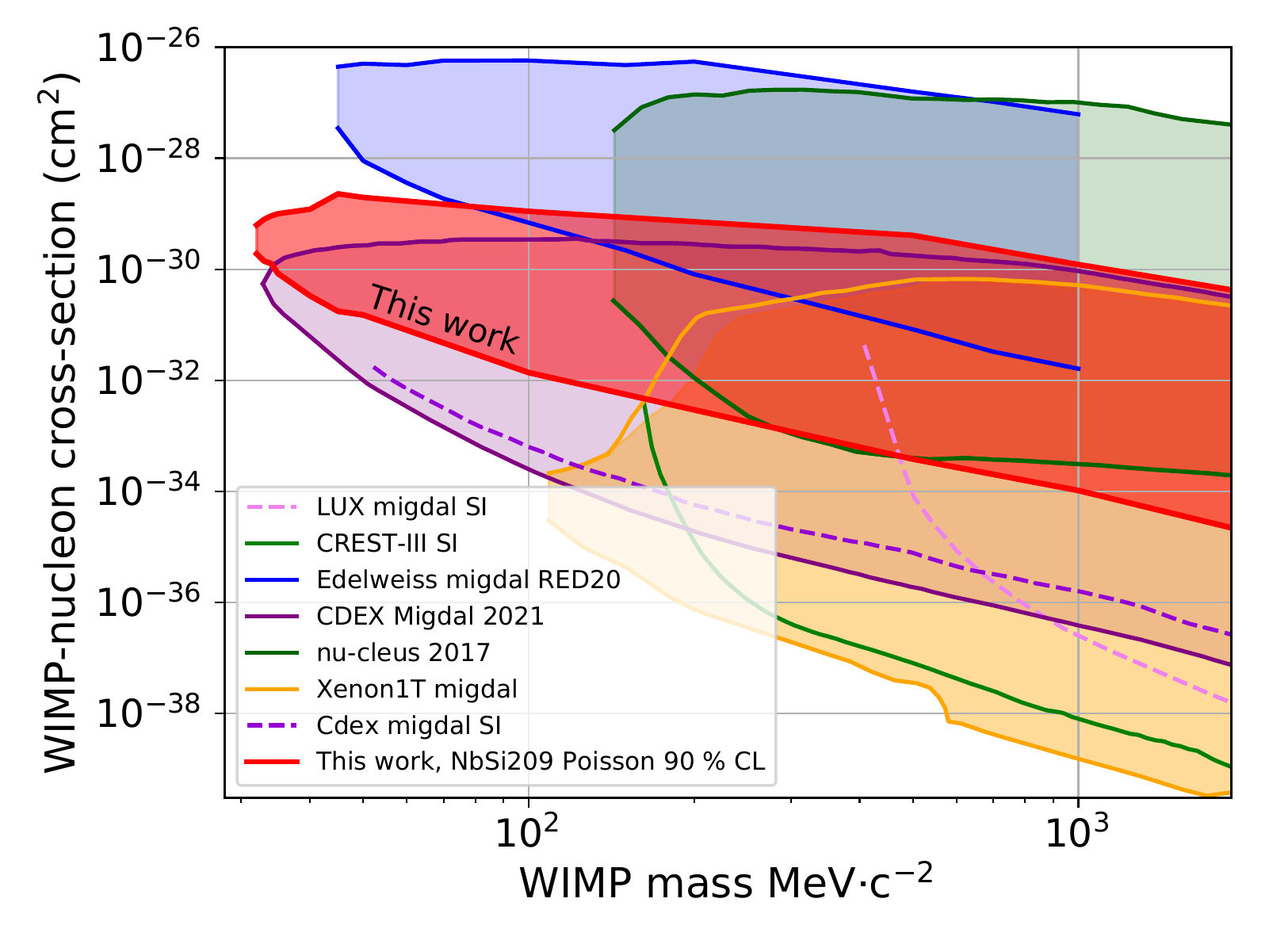}
\caption{$90\%$~C.L. upper limit on the cross-section for Spin-Independent interaction  between DM and Ge nuclei through the Migdal effect. The red contour is obtained by taking into account the slowing of the DM particle flux through the material above the detector. These results are compared to other experiments \cite{Akerib:2018hck,nucleus,cresst3,xenon-migdal,cedex,cedex2021} (see text). 
}\label{limit_FINAL}
\end{center}

\end{figure}

The  $90 \%$~C.L. excluded region presented in Fig.~\ref{figeesevsstd} constrains masses down to $32$~MeV$\cdot$c$^{-2}$. 
Below this value of $32$~MeV$\cdot$c$^{-2}$, the very large cross-section needed to yield an observable signal leads to stopping effect from overburden and shielding, consequently, no constraint can be obtained for lower masses in this analysis.
This contour is shown in Fig.~\ref{limit_FINAL} along with other experimental results \cite{Akerib:2018hck,nucleus,cresst3,xenon-migdal,cedex,cedex2021}. It constrains a new region of parameter space for cross-sections close to $10^{-29}$~cm$^2$ and masses between $32$ and $100$~MeV$\cdot$c$^{-2}$. 
This contour is also compared to the previous results of an EDELWEISS-surf Midgal search~\cite{red20} which was performed at the surface. Orders of magnitude of improvement have been achieved. The underground operation did not jeopardize the potential of this search, despite the enhanced Earth-shielding from the larger overburden, thanks to the significant reduction of the background level obtained in the EDELWEISS-III setup at LSM. 
The effective threshold of $30$~eV$_{ee}$ achieved here, lower than the $60$~eV threshold of Ref.~\cite{red20}, contributes to the extension of the search to masses below $45$~MeV$\cdot$c$^{-2}$.
This threshold is more than five times lower than those of CDEX~\cite{cedex2021} ($160$~eV$_{ee}$) and XENON~\cite{xenon-migdal} ($\sim 200$~eV$_{ee}$). 
However, both experiments achieved significantly better background levels, and this aspect is clearly the main limiting factor for the use of the present detector to probe lower cross-sections.

\section{Discussion}\label{sec:discussion}

\begin{figure}[!h]
\begin{center}
\includegraphics[width=0.99\linewidth]{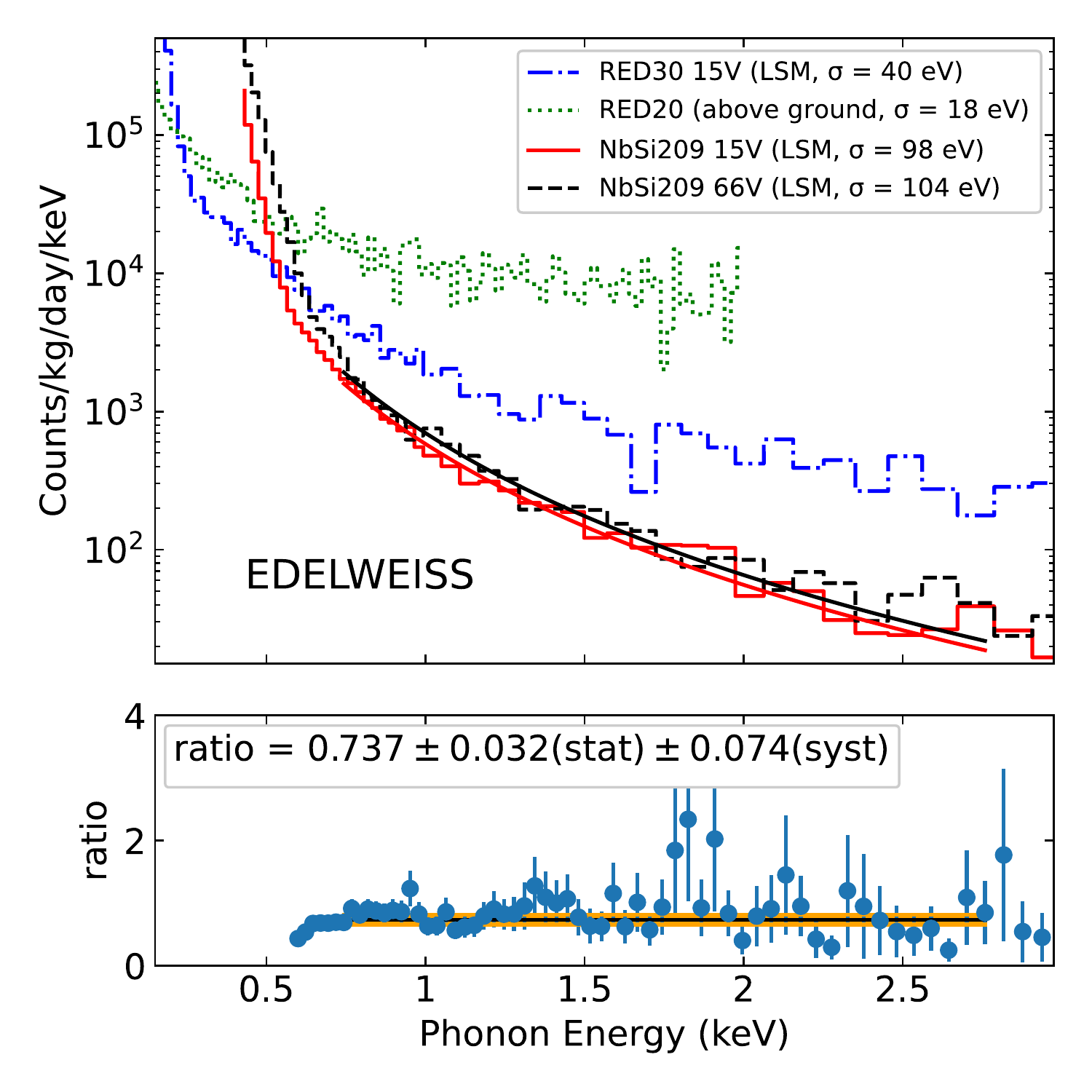}
\caption{Heat energy spectra of events recorded with RED20 operated above ground at $0$~V (green)\cite{red20}, events with  no ionization ($E_{ion} < 0$) for the RED30 detector operated at $15$~V (blue),  NbSi209 operated at $15$~V (red) and $66$~V (black). The fitted power law ($\alpha E ^{\beta}$) on NbSi spectra when operated at  $66$~V and $15$~V in black ($\alpha = 697.9 \pm 16.8$ ; $\beta  = -3.41 \pm 0.08$) and red ($\alpha = 582.0 \pm 11.2$ ; $\beta  = -3.39 \pm 0.07$) respectively. The spectra are corrected for efficiency, assuming heat-only events for RED30 and NbSi209. The lower figure shows the ratio of NbSi209 distributions recorded at $15$ and $66$~V with its statistical uncertainties (blue) and the associated fit of a constant (black line) and its uncertainty (statistical and systematic) band  (orange).}\label{data_graph}
\end{center}
\end{figure}

As the search for DM particles appears to be limited by backgrounds, these were investigated more thoroughly.
It was found that most of the population in the energy interval between $0.8$ to $3$~keV come from events where a heat energy deposit is not associated with any ionization, since they are not affected by the NTL amplification. 
This is shown in Fig.~\ref{data_graph} that compares the data recorded by NbSi209 at biases of $15$~V and $66$~V.
The rise below $600$~eV is mainly due to the read-out noise, which slightly increases at $66$~V.
The compatibility of the two spectra above $0.8$~keV indicates that most events in that region are not affected by the NTL amplification.
The fit of a power law ($\alpha E ^ {\beta}$) yields identical slopes within uncertainties $\beta \sim 3.40 $ for both spectra.
This is further illustrated by the flatness of the ratio of the two spectra (bottom panel of Fig.~\ref{data_graph}) as a function of energy.
The value of this ratio is $0.74 \pm 0.03 ~(stat) \pm 0.07 ~(syst)$,
where the central value is fitted in the range from $0.8$ to $2.8$~keV, and the systematic error considers variations of this range.
This ratio depends on the fraction $x$ of events associated with charges.
It should be equal to one for $x=0$ in the absence of time-dependence of the rates \cite{excess}.
The observed ratio is compatible with values deduced from the long-term variations observed at 15V, but the deviation from 1 will be taken as a conservative systematic uncertainty associated to the time variations.
Assuming a worst-case scenario where the HO and ER populations follow the same power spectrum, the resulting upper limit is $x<0.0004$ at 90\%C.L.
This confirms that HO events dominate the spectra in the range from $0.8$ to $2.8$~keV.

At $66$~V, this diagnostic concerns events with electron-equivalent energies between $25$ and $130$~eV$_{ee}$, well below what can be probed by the ionization signal resolution of $210$~eV$_{ee}$.
In these spectra, the cut on $E_{ion}> 400$~eV$_{ee}$ has been replaced by a cut on $E_{ion}< 0$~eV$_{ee}$ in order to accurately measure the contribution of HO events.
Considering that the contribution of events with $E_{ion}< 0$~eV$_{ee}$ does not affect significantly the efficiency-corrected rates below $3$~keV, 
it is a further indication that the HO population dominates the low-energy region.

The background rate after efficiency correction is $\sim 600 $ events.kg$^{-1}$.day$^{-1}$.keV$^{-1}$ at 1 keV, corresponding to $1.4\times 10^4$~events.kg$^{-1}$.day$^{-1}$.keV$_{ee}^{-1}$ at $43$~eV$_{ee}$ for the 66~V data.
For comparison, Fig.~\ref{data_graph} also shows the spectra observed in an EDELWEISS-surf $34$~g detector equipped with Ge-NTD heat sensors, operated above-ground at $0$~V~\cite{red20} and at LSM at $15$~V~\cite{red30}.
In ~\cite{red30,ltd19}, 
it was established by comparing data recorded at $15$~V and $78$~V that at LSM the $34$~g detector spectrum was dominated by the HO background.
In the energy range relevant for Migdal searches, i.e. above $30$~eV$_{ee}$ (690~eV), the backgrounds in NbSi209 are lower than those of Ref.~\cite{red30}, showing that the present detector is better suited for this type of searches, in terms of HO event rates. 
The green line shows the spectrum of the EDELWEISS-surf experiment \cite{red20}. The comparatively higher rate highlights the reduction of backgrounds achieved underground in the EDELWEISS-III setup and the consequent gain in sensitivity.

Studies to better understand the origin of these still unexplained HO events are ongoing.
This background not only affects EDELWEISS detector but is a limiting factor for numerous experiments in the direct DM search cryogenic community \cite{excess}. 
Although the HO event rate per unit mass appears to be reduced relative to those observed in a smaller detector equipped with an NTD sensor, the change of sensor technology (with different sensitivities to thermal and out-of-equilibrium phonons) does not eliminate this type of events entirely.
This excludes, for example, stress due to the gluing of the Ge-NTD as being the dominant cause of HO events.


\section{Conclusion} \label{sec:conclusions}
The EDELWEISS collaboration has searched for DM particle interaction exploiting the Migdal effect with masses between $32$~MeV$\cdot$c$^{-2}$ and $2$~GeV$\cdot$c$^{-2}$ using a $200$~g Ge detector operated underground at the Laboratoire Souterrain de Modane. 
The phonon signal was read out using a Transition Edge Sensor made of a NbSi thin film.
The detector was biased at $66$~V in order to benefit from NTL amplification and resulting in a resolution on the energy of electron recoils of $4.46$~eV$_{ee}$ (RMS). 
The effective analysis threshold of $30$~eV$_{ee}$ is better than other Migdal searches, but is limited by a large background not associated to charge creation in the detector, whose cause remains to be identified. 
The search constrains a new region of parameter space 
for cross-sections close to $10^{-29}$~cm$^2$ and masses
between $32$ and $100$~MeV$\cdot$c$^{-2}$.
The reduction of threshold achieved with the NbSi sensor shows the relevance of its use for out-of-equilibrium phonon sensitive devices for low-mass DM searches. 
In the context of its EDELWEISS-SubGeV program, the collaboration is also investigating new methods to significantly reduce HO backgrounds by improving its ionization resolution with the use of new cold preamplifiers~\cite{hemt}, and by developing NbSi-instrumented devices able to tag the out-of-equilibrium NTL phonons associated to a single electron.

\begin{acknowledgments}

The EDELWEISS project is supported in part by the French Agence Nationale pour la Recherche (ANR) and the LabEx Lyon Institute of Origins (ANR-10-LABX-0066) of the Universit\'e de Lyon within the program ``Investissements d'Avenir'' (ANR-11-IDEX-00007), by the P2IO LabEx (ANR-10-LABX-0038) in the framework ``Investissements d'Avenir'' (ANR-11-IDEX-0003-01) managed by the ANR (France), and the Russian Foundation for Basic Research (grant No. 18-02-00159). This project  has  received  funding  from  the  European Union’s Horizon 2020 research and innovation program under the Marie Sk\l odowska-Curie Grant Agreement No. 838537. B.J. Kavanagh thanks the Spanish Agencia Estatal de Investigaci\'on (AEI, MICIU) for the support to the Unidad de Excelencia Mar\'ia de Maeztu Instituto de F\'isica de Cantabria, ref. MDM-2017-0765.
We thank J.P. Lopez (IP2I) and the Physics Department of Universit\'{e} Lyon 1 for their contribution to the radioactive sources.

\end{acknowledgments}

\appendix*

\newcommand{\arXiv}[1]{\href{https://arxiv.org/abs/#1}{arXiv:#1}}
\newcommand{\oldarXiv}[1]{\href{https://arxiv.org/abs/#1}{#1}}
\newcommand{\DOI}{https://doi.org}


\frenchspacing


\begin{thebibliography}{100}

\bibitem{Goodman:1984dc}
  Goodman, Mark W. and Witten, Edward
  \href{\DOI/10.1103/PhysRevD.31.3059}{Phys.\ Rev.\ D.\ {\bf 31}, 3059 (1985) }. 

\bibitem{Drukier:1986tm}
A. Drukier, K. Freese, D. N. Spergel
\href{\DOI/10.1103/PhysRevD.33.3495}{Phys.\ Rev.\  D.\ {\bf33}, 3495-3508 (1986).}

\bibitem{Drukier:1984vhf} 
A. Drukier, L. Stodolsky,  
 \href{\DOI/10.1103/PhysRevD.30.2295}{Phys.\ Rev.\ D.\ {\bf 30}, 2295 (1984)}. 

\bibitem{xenon1t} 
E. Aprile {\it et al.} (XENON Collaboration),
\href{\DOI/10.1103/PhysRevLett.121.111302}{Phys.\ Rev.\ Lett.\ {\bf 121}, 111302 (2018)}, \arXiv{1805.12562}.

\bibitem{lux} 
D. S. Akerib {\it et al.} (LUX Collaboration),
\href{\DOI/10.1103/PhysRevLett.118.021303}{Phys.\ Rev.\ Lett.\ {\bf 118}, 021303 (2017)}, \arXiv{1608.07648}.

\bibitem{pandax} 
A. Tan {\it et al.} (PandaX-II Collaboration),
\href{\DOI/10.1103/PhysRevLett.117.121303}{Phys.\ Rev.\ Lett.\ \textbf{117}, 121303 (2016)}, \arXiv{1607.07400}.

\bibitem{Essig} 
R. Essig, J. Kaplan, P. Schuster and N. Toro,  (2010),
\arXiv{1004.0691}.

\bibitem{Cheung} 
C. Cheung, J. T. Ruderman, L.-T. Wang and I. Yavin,
\href{\DOI/10.1103/PhysRevD.80.035008}{Phys.\ Rev.\ D {\bf 80}, 035008 (2009)}, \arXiv{0902.3246}.

\bibitem{Hooper} 
D. Hooper and W. Xue, 
\href{\DOI/10.1103/PhysRevLett.110.041302}{Phys.\ Rev.\ Lett.\ {\bf 110}, 041302 (2013)}, \arXiv{1210.1220}.

\bibitem{Falkowski} 
A. Falkowski, J.T. Ruderman and T. Volansky,
\href{\DOI/10.1007/JHEP05(2011)106}{JHEP {\bf 1105}, 106 (2011)}, \arXiv{1101.4936}.

\bibitem{Petraki} 
K. Petraki and R.R. Volkas, 
\href{\DOI/10.1142/S0217751X13300287}{Int. J. Mod. Phys. A {\bf 28}, 1330028 (2013)}, \arXiv{1305.4939}.

\bibitem{Zurek} 
K. M. Zurek, 
\href{\DOI/10.1016/j.physrep.2013.12.001}{Phys.\ Rep.\ {\bf 537}, 91 (2014)}, \arXiv{1308.0338}.

\bibitem{Bertone:2018krk} 
G.~Bertone and T.~M.~P.~Tait,
  \href{\DOI/10.1038/s41586-018-0542-z}{Nature {\bf 562}, 51-56 (2018)}, \arXiv{1810.01668}.

\bibitem{red30}
Q. Arnaud {\it et al.} (EDELWEISS Collaboration),
\href{10.1103/physrevlett.125.141301}{Phys.\ Rev.\ Lett.\ \textbf{135}, 141301 (2020)},
\arXiv{2003.01046}


\bibitem{cresst} 
G. Angloher {\it et al.} (CRESST Collaboration), \href{\DOI/10.1140/epjc/s10052-016-3877-3}{Eur.\ Phys.\ J.\ C.\ {\bf 76}, 25 (2016)}, \arXiv{1509.01515}.

\bibitem{cdmslite} 
R. Agnese {\it et al.} (SuperCDMS Collaboration),
\href{\DOI/10.1103/PhysRevD.97.022002}{Phys.\ Rev.\ D {\bf 97}, 022002 (2018)}, \arXiv{1707.01632}.

\bibitem{damic}
A. Aguilar-Arevalo {\it et al.} (DAMIC Collaboration),
\href{10.1103/PhysRevLett.118.141803}{Phys.\ Rev.\ Lett.\ {\bf 118}, 141803 (2017)}, \arXiv{1611.03066}.

\bibitem{sensei}
O. Abramoff {\it et al.} (SENSEI Collaboration),
\href{10.1103/physrevlett.122.161801}{Phys.\ Rev.\ Lett.\ {\bf 122}, 161801 (2019)}, \arXiv{2004.11378}.

\bibitem{cedex}
Z.Z. Liu {\it et al.} (CDEX Collaboration),
\href{10.1103/physrevlett.123.161301}{Phys.\ Rev.\ Lett.\ {\bf 123}, 161301 (2019)} \arXiv{1905.00354}.

\bibitem{nucleus} 
G.~Angloher {\it et al.} (CRESST Collaboration),
\href{\DOI/10.1140/epjc/s10052-017-5223-9}{Eur.\ Phys.\ J.\ C {\bf 77}, 637 (2017)}, \arXiv{1707.06749}.

\bibitem{SuperCDMS:2020aus}
D. W. Amaral {\it et al.}
\href{\DOI/10.1103/PhysRevD.102.091101}{Phys.\ Rev.\ D.\ {\bf 102}, 091101(2020)}

\bibitem{red20}
E. Armengaud {\it et al.} (EDELWEISS Collaboration),
\href{10.1103/physrevd.99.082003}{Phys.\ Rev.\ D \textbf{99}, 082003 (2019)},
\arXiv{1901.03588}

\bibitem{cedex2021}
Z.Z. Liu {\it et al.} (CEDEX Collaboration),
\arXiv{arXiv:2111.11243}.
 
\bibitem{Vergados:2004bm} 
J.~D.~Vergados and H.~Ejiri,
\href{\DOI/10.1016/j.physletb.2004.11.085}{Phys.\ Lett.\ B {\bf 606}, 313 (2005)}, \oldarXiv{hep-ph/0401151}.

\bibitem{Moustakidis:2005gx} 
C.~C.~Moustakidis, J.~D.~Vergados and H.~Ejiri,
\href{\DOI/10.1016/j.nuclphysb.2005.08.033}{Nucl.\ Phys.\ B {\bf 727}, 406 (2005)}, \oldarXiv{hep-ph/0507123}.

\bibitem{Bernabei:2007jz} 
R.~Bernabei {\it et al.},
  \href{\DOI/10.1142/S0217751X07037093}{Int.\ J.\ Mod.\ Phys.\ A {\bf 22}, 3155 (2007)}, \arXiv{0706.1421}.

\bibitem{Ibe:2017yqa} 
M.~Ibe, W.~Nakano, Y.~Shoji and K.~Suzuki,
  \href{\DOI/10.1007/JHEP03(2018)194}{JHEP {\bf 1803}, 194 (2018)}, \arXiv{1707.07258}.

\bibitem{Dolan:2017xbu} 
M.~J.~Dolan, F.~Kahlhoefer and C.~McCabe,
  \href{\DOI/10.1103/PhysRevLett.121.101801}{Phys.\ Rev.\ Lett.\  {\bf 121}, 101801 (2018)}, \arXiv{1711.09906}.

\bibitem{Akerib:2018hck} 
D.~S.~Akerib {\it et al.} (LUX Collaboration),\href{\DOI/10.1103/physrevlett.122.131301}{Phys.\ Rev.\ Lett.\ {\bf 122}, 131301 (2019)} \arXiv{1811.11241}.

\bibitem{xenon-migdal}
E. Aprile {\it et al.} (XENON Collaboration),
\href{\DOI/10.1103/PhysRevLett.123.241803}{Phys.\ Rev.\ Lett.\ {\bf 123}, 241803}, \arXiv{1907.12771}.

\bibitem{outer-essig}
R. Essig, J. Pradler, M. Sholapurkar and T.-T. Yu,
\href{10.1103/physrevlett.124.021801}{Phys.\ Rev.\ Lett.\  {\bf 124}, 021801 (2020)}, \arXiv{1908.10881}.

\bibitem{outer-liu}
C.-P. Liu, C.-P. Wu, H.-C. Chi and J.-W. Chen,
\href{10.1103/physrevd.102.121303}{Phys.\ Rev.\ D  {\bf 102}, 121303 (2020)}, \arXiv{arXiv:2007.10965}.

\bibitem{edwtech} 
E. Armengaud {\it et al.} (EDELWEISS Collaboration), 
\href{\DOI/10.1088/1748-0221/12/08/P08010}{JINST {\bf 12}, P08010 (2017)}, \arXiv{1706.01070}.

\bibitem{nbsi-ltd}
S. Marnieros {\it et al.} (EDELWEISS Collaboration), 
Submitted to JLTP, Special Issue for the 19th International Workshop on Low Temperature Detectors, \arXiv{2201.01639}.

\bibitem{Neganov} 
B. Neganov and V. Trofimov, Otkryt.\ Izobret.\ {\bf 146}, 215 (1985), USSR Patent No. 1037771.

\bibitem{Luke} 
P. N. Luke, J.\ Appl.\ Phys.\ {\bf 64}, 6858 (1988).

\bibitem{knoll}
G. F. Knoll, Radiation Detection and Measurement, 4th ed.
(John Wiley and Sons, New York, 2010).

\bibitem{trigger} 
S.~Di Domizio, F.~Orio and M.~Vignati,
\href{\DOI/10.1088/1748-0221/6/02/P02007}{JINST {\bf 6}, P02007 (2011)}, \arXiv{1012.1263}.

\bibitem{2021semicondmigdal} 
Knapen, Simon and Kozaczuk, Jonathan and Lin, Tongyan,
\href{\DOI/10.1103/physrevlett.127.081805}{Phys.\ Rev.\ Lett.\  {\bf 127}, no. 8, 081805 (2021)}.

\bibitem{Essig:2019xkx}
Essig,  Pradler,  Sholapurkar and Yu,
\href{\DOI/10.1103/physrevlett.124.021801}{Phys.\ Rev.\ Lett.\ {\bf 124}, 021801 (2020)}

\bibitem{Hochberg:2021pkt}
Y. Hochberg, E. D. Kramer, N. Kurinsky, B. V. Lehmann
\arXiv{2109.04473}

\bibitem{standardassumptions} 
J.~D.~Lewin and P.~F.~Smith, 
\href{\DOI/10.1016/S0927-6505(96)00047-3}{Astropart.\ Phys.\ {\bf 6}, 87 (1996)}

\bibitem{Helm:1956zz}
Richard H. Helm
\href{\DOI/10.1103/PhysRev.104.1466}{Phys.\ Rev.\ {\bf 104}, 1466 (1956)}

\bibitem{Digman:2019wdm}
M. C. Digman, C. V. Cappiello, J. F. Beacom, C. M. Hirata, A. H. G. Pete
\href{\DOI/10.1103/PhysRevD.100.063013}{Phys.\ Rev.\ D {\bf 100}, 063013 (2019)}

\bibitem{quenchingpaper}
B.J. Scholz, A.E. Chavarria, J.I. Collar, P. Privitera, A.E. Robinson
\href{\DOI/10.1103/PhysRevD.94.122003}{Phys.\ Rev.\ D {\bf 94}, 122003 (2016)}.


\bibitem{Kavanagh:2016pyr}
B. J. Kavanagh, R. Catena and C. Kouvaris, 
\href{\DOI/10.1088/1475-7516/2017/01/012}{JCAP {\bf 1701} (2017) 012.}

\bibitem{Emken:2017qmp}
T. Emken, C. Kouvaris. \href{\DOI/10.1088/1475-7516/2017/10/031}{JCAP {\bf 10} (2017) 031}

\bibitem{Mahdawi:2018euy}
M. Shafi Mahdawi, Glennys R. Farrar, \arXiv{1712.01170}

\bibitem{Hooper:2018bfw}
D. Hooper and S. D. McDermott,
\href{\DOI/10.1103/PhysRevD.97.115006}{Phys. Rev. D {\bf 97}, 115006 (2018)}
\arXiv{1802.03025}

\bibitem{verne} 
B.~J.~Kavanagh, ``verne v1.0 [computer software]'', Astrophysics  Source  Code  Library,  record \href{http://ascl.net/1802.005}{ascl:1802.005} (archived on Zenodo, \href{\DOI/10.5281/zenodo.1115601}{DOI:10.5281/zenodo.1115601 (2017)}) .

\bibitem{Kavanagh:2017cru} 
B.~J.~Kavanagh,
\href{\DOI/10.1103/PhysRevD.97.123013}{Phys.\ Rev.\ D {\bf 97}, 123013 (2018)}, \arXiv{1712.04901}.
  
\bibitem{Starkman:1990nj} 
G.~D.~Starkman, A.~Gould, R.~Esmailzadeh and S.~Dimopoulos,
\href{\DOI/10.1103/PhysRevD.41.3594}{Phys.\ Rev.\ D {\bf 41}, 3594 (1990)}.
  
\bibitem{Emken:2018run} 
T.~Emken, C.~Kouvaris,
\href{\DOI/10.1103/PhysRevD.97.115047}{Phys.\ Rev.\ D {\bf 97}, 115047 (2018)}, \arXiv{1802.04764}.

\bibitem{Gould:1987ww} 
A.~Gould,
\href{\DOI/10.1086/166347}{Astrophys.\ J.\  {\bf 328}, 919 (1988)}.

\bibitem{Mack:2007xj} 
G.~D.~Mack, J.~F.~Beacom and G.~Bertone,
\href{\DOI/10.1103/PhysRevD.76.043523}{Phys.\ Rev.\ D {\bf 76}, 043523 (2007)}, \arXiv{0705.4298}.

\bibitem{Kouvaris:2015xga} 
C.~Kouvaris and N.~G.~Nielsen,
\href{\DOI/10.1103/PhysRevD.92.075016}{Phys.\ Rev.\ D {\bf 92}, 075016 (2015)}, \arXiv{1505.02615}.

\bibitem{cresst3} 
A. Abdelhameed {\it et al}. (CRESST),
\href{\DOI/10.1103/PhysRevD.100.102002}	{Phys. Rev. D {\bf 100}, 102002 (2019)}
\arXiv{1904.00498}.

\bibitem{ltd19}
J. Gascon {\it et al}, Submitted to JLTP, Special Issue for the 19th International Workshop on Low Temperature Detectors, \arXiv{2112.05467}.

\bibitem{excess}
P. Adari {\it et al}, \arXiv{2202.05097}


\bibitem{hemt} 
A. Juillard {\it et al}, \href{\DOI/10.1007/s10909-019-02269-5}{J. Low Temp. Phys. {\bf 199}, 798 (2020)}.









































%

















  





  
  
  
  
  
  








   


  





      


\end{thebibliography}
\end{document}